\documentclass[aps,prd,onecolumn,nofootinbib,superscriptaddress]{revtex4-2}

\usepackage{amsmath,amssymb,amsfonts}
\usepackage{mathtools, physics2, commath}
\usepackage{graphicx}
\usepackage{hyperref}
\usepackage{xcolor}
\usepackage{nicefrac}

\newcommand{\m}{\mu}
\newcommand{\n}{\nu}

\newcommand{\f}{\phi}

\newcommand{\s}{\sigma}

\newcommand{\vf}{\varphi}

\newcommand{\nn}{\nonumber}
\newcommand{\sq}{\sqrt}
\newcommand{\sqdet}{\sq{-g}}

\newcommand{\cL}{\mathcal{L}}
\newcommand{\cH}{\mathcal{H}}

\newcommand{\der}{\nabla}

\newcommand{\diff}{\mathrm{d}}

\newcommand{\cLm}{\mathcal{L}_m}

\newcommand{\cG}{\mathcal{G}}
\newcommand{\cLM}{\cL^{m}_{red}}
\newcommand{\cLG}{\cL^{g}_{red}}
\newcommand{\Dz}{\mathcal{D}_z}
\newcommand{\Zms}{ \mathcal{Z}_{\mathrm{ms}}}\newcommand{\cLN}{\cL^{m,N}_{red}}

\begin{document}
	
	\title{Minisuperspace Double Copy in Lifshitz Spacetimes}
	
	\author{Mehmet Kemal G\"um\"u\c{s}}
	\email{gumus@ijclab.in2p3.fr}
	\altaffiliation[Present address: ]{UMR 9012, Laboratoire de Physique des 2 Infinis Ir\`ene Joliot-Curie (IJCLab), CNRS, Universit\'e Paris-Saclay, 91405 Orsay, France.}
	\affiliation{Independent Researcher, 1929. Sokak 2-A/20, Yenimahalle, 06370 Ankara, Turkey}
	
	\date{\today}
	
\begin{abstract}
We develop a minisuperspace formulation of the classical double copy for static Lifshitz spacetimes in arbitrary dimension. The reduced gravitational action takes a universal total-derivative form in which all theory dependence is captured by a single potential. A weighted radial operator acting on the reduced gravitational Lagrangian reproduces, off shell, the Maxwell operator for the temporal component of the single-copy field, up to a residual term that encodes anisotropic scaling and horizon geometry and vanishes in the relativistic limit. After imposing the Hamiltonian constraint, the matter sector generates the corresponding source, reproducing known single-copy charge densities when a Kerr--Schild description exists. The same mechanism persists in Lovelock-type theories, where the Kerr--Schild and Weyl constructions are not available. We illustrate the construction with Lifshitz black holes of different matter content and horizon topology, and with a charged AdS solution of Einstein--Gauss--Bonnet gravity.
\end{abstract}
	\maketitle

\section{Introduction}

The double copy has revealed a remarkable relation between gauge theory and gravity. 
It first appeared in perturbative scattering amplitudes, where gravity amplitudes can be obtained from gauge-theory amplitudes by replacing color factors with kinematic ones \cite{Bern:2008qj}. 
This raised a broader question: is the double copy only a special feature of perturbation theory, or does it reflect a deeper relation between gauge and gravitational dynamics?

At the classical level, this question has been studied most successfully in two main ways. 
The first is the Kerr--Schild double copy (KSDC). In this framework, some metrics can be written in a form that makes the Einstein equations effectively linear in a scalar function \cite{Monteiro:2014cda}. 
When this happens, one can associate the gravitational solution with a Maxwell field and obtain an exact classical double copy. 
The second is the Weyl double copy (WDC), which relates algebraically special gravitational fields to gauge fields through spinorial objects \cite{Luna:2018dpt,Godazgar:2020zbv}. 
These approaches have greatly advanced the classical double-copy program and clarified many of its geometric aspects \cite{CarrilloGonzalez:2017iyj,Easson:2021asd,Easson:2022zoh,Alkac:2023glx}. 
The literature has also expanded quickly in recent years, including work on curved backgrounds, alternative ans\"atze, algebraically general geometries, and further developments of both KSDC and WDC (see, e.g., Refs.~\cite{Monteiro:2014cda,CarrilloGonzalez:2017iyj,Luna:2018dpt,Godazgar:2020zbv,Easson:2021asd,Easson:2022zoh,Alkac:2023glx,Alkac:2025iyw, Alencar:2026zdz, Armstrong-Williams:2024bog, Easson:2023dbk, Chawla:2023bsu, Adamo:2022dcm, Alkac:2021seh, Alkac:2022tvc, Gumus:2020hbb}). 

Recent work has also pushed the double-copy idea beyond Einstein gravity. This includes higher-curvature theories such as quasitopological gravity, where part of the correspondence may survive in suitably reduced non-Einstein settings \cite{Frolov:2025ddw}. At the same time, these developments highlight an important limitation: most known constructions rely on special coordinates, special algebraic properties, or specially chosen ans\"atze.

This motivates a different point of view. Instead of asking whether a spacetime admits a particular geometric decomposition, one may ask whether double-copy structure is already visible in the \emph{reduced dynamics} of gravity within a highly symmetric sector. A recent step in this direction was taken in \cite{Alkac:2024pfd}. That work formulated the classical double copy in a minisuperspace setting for static, spherically symmetric black holes, deriving the correspondence directly from the reduced action. Within this framework, the gravitational system reduces to an effective one-dimensional radial problem, and a Maxwell-like structure emerges at the level of the reduced dynamics. Importantly, the construction extends beyond Einstein gravity to Lovelock and quasi-topological theories, where the single-copy potential becomes a polynomial function of the Kerr--Schild (KS) scalar rather than matching it directly. This suggests that part of the classical double copy may be encoded not only in special solutions, but also in the constraint structure of symmetry-reduced gravity, and that this structure can persist in higher-curvature settings.

In this paper, we develop this idea for \emph{Lifshitz spacetimes}, which are defined by the anisotropic scaling
\begin{equation}
	t\to \lambda^z t,
	\qquad
	x^i\to \lambda x^i.
\end{equation}
These geometries play an important role in non-relativistic holography and in gravitational models with anisotropic scaling. They also provide a harder test for the classical double copy than anti-de Sitter (AdS) space. Unlike AdS, Lifshitz spacetimes are not maximally symmetric, and they usually require matter fields to support the geometry. For this reason, they offer a natural setting in which to ask whether a double-copy-like structure survives away from constant-curvature backgrounds.

Our approach provides an alternative perspective on KSDC. Rather than starting from a special metric decomposition, we extract the single-copy structure directly from the \emph{reduced action}. More precisely, we show that in a minisuperspace reduction of static Lifshitz spacetimes, the reduced gravitational sector takes a universal form. From this form, one can identify the radial Maxwell operator \emph{off shell} as a kinematical feature of the reduced system. The corresponding source term appears only \emph{on shell}, through the radial Hamiltonian constraint, and is fixed by the matter sector together with the departure from maximal symmetry. In this way, the double copy is recast as a property of reduced gravitational dynamics rather than as a consequence of a special geometric ansatz.

To implement this program, we begin with a static metric ansatz compatible with Lifshitz asymptotics and depending only on the radial coordinate. We do not fix the Lifshitz scaling form at the start. Instead, we first carry out the minisuperspace reduction with an arbitrary radial lapse function, and only then specialize to the lapse adapted to Lifshitz scaling. This order is important. It preserves the radial constraint structure of the theory and makes it clear which parts of the construction are kinematical and which are solution-dependent.

Our main result is that the reduced gravitational action naturally produces the radial operator that governs the time component of Maxwell's equations for an effective electric potential. In other words, the Maxwell operator does not need to be introduced by hand. It emerges directly from the reduced gravitational dynamics. For Lifshitz backgrounds, however, the result is not identical to the one found in maximally symmetric AdS. An additional term appears because Lifshitz spacetime is not maximally symmetric. We show that this extra term has a universal structure: one part is tied to anisotropic scaling, while the other is fixed by the curvature of the horizon geometry. In the relativistic limit $z\to1$, this term vanishes smoothly and the AdS behavior is recovered.

Once the reduced equations of motion are imposed, the same framework gives a sourced Maxwell equation. In this on-shell form, the matter sector determines the effective charge density. Whenever a KSDC description exists, the source obtained in minisuperspace reproduces the corresponding KSDC single-copy result. More generally, the reduced formalism separates two logically distinct ingredients: an off-shell operator identity coming from the universal structure of the reduced gravitational action, and an on-shell source determined by matter content and by the departure from maximal symmetry.

An additional advantage of this method is that it extends naturally to higher-curvature theories. For Lovelock gravities, the reduced action is again controlled by a single effective radial function, now built from the corresponding Wheeler polynomial \cite{Wheeler:1985nh,Wheeler:1986zp}. Since the standard Kerr--Schild and Weyl double copies are formulated for Einstein gravity (first-order Lovelock), this extension goes beyond them. We show that the same reduced-action mechanism still produces the Maxwell operator, together with the same universal deviation structure. This suggests that the minisuperspace double copy is not peculiar to Einstein gravity, but instead reflects a broader feature of symmetry-reduced gravitational dynamics.

To make the discussion concrete, we study several explicit examples of Lifshitz black holes. These include a case with an effectively vacuum single copy in KSDC, a charged Einstein--Proca example with a non-trivial source, and a topological Lifshitz black hole in which horizon curvature plays an essential role. As a relativistic consistency check, we also study the charged AdS solution of Einstein--Gauss--Bonnet gravity. Although this final example is not Lifshitz, it provides a clean test of the higher-curvature extension in a setting where an exact solution is known.
	
The paper is organized as follows. 
In Sec.~\ref{KSDC} we briefly review the Kerr--Schild double copy for static Lifshitz black holes and summarize the corresponding single- and zeroth-copy equations. 
In Sec.~\ref{MDC} we perform the minisuperspace reduction and derive the reduced Maxwell structure, first in Einstein gravity and then in Lovelock theories. 
In Sec.~\ref{static_spherical} we apply the formalism to the simplest setting, namely a static, spherically symmetric black hole.
In Sec.~\ref{LifshitzExamples} we examine explicit black hole examples and compare the resulting source terms with the Kerr--Schild picture. 
We conclude in Sec.~\ref{Conc} with a discussion of the broader implications of the minisuperspace perspective for the classical double copy.

	\section{Kerr--Schild Double Copy in Lifshitz Spacetimes} \label{KSDC}
	
	The KSDC formalism begins with the trace-reversed Einstein equations written with mixed indices,
	\begin{equation}
		{R^\mu}_\n-\frac{2\Lambda}{d-2}\,\delta^\mu_{\ \nu}
		=\widetilde{T}^\mu_{\ \, \nu},
		\label{reversed}
	\end{equation}
	where $\Lambda$ is the cosmological constant and the trace-reversed energy-momentum tensor is defined as
	\begin{equation}
		\widetilde{T}^\mu_{\ \nu}
		=
	{	T^\mu}_{\nu}
		-\frac{1}{d-2}\,{\delta^\mu}_{\nu}\,T,
		\qquad
		T={T^\mu}_{\mu}.
	\end{equation}
	This formulation is particularly suitable for the Kerr--Schild construction, since the Ricci tensor with mixed indices becomes linear in the metric perturbation.
	
	Following \cite{Alkac:2021bav}, we do not assume any special property of the background metric. Instead, we assume that the spacetime admits Kerr--Schild coordinates in which the metric can be written as
	\begin{equation}
		g_{\mu\nu}
		=
		\bar g_{\mu\nu}
		+
		\phi_{KS} \, k_\mu k_\nu,
		\label{KS}
	\end{equation}
	where $\bar g_{\mu\nu}$ is the background metric, $\phi_{KS}$ is the Kerr--Schild scalar, and $k_\mu$ is a null and geodesic vector with respect to both the background and the full metric.
	
	For Kerr--Schild geometries, the mixed Ricci tensor takes the form \cite{Stephani:2003tm,CarrilloGonzalez:2017iyj}
	\begin{equation}
		R^\mu_{\ \nu}
		=
		\bar R^\mu_{\ \nu}
		-\phi_{KS}\, k^\mu k^\alpha \bar R_{\alpha\nu}
		+\frac{1}{2}\Big[
		\bar\nabla^\alpha \bar\nabla^\mu(\phi_{KS} k_\alpha k_\nu)
		+\bar\nabla^\alpha \bar\nabla_\nu(\phi_{KS} k^\mu k_\alpha)
		-\bar\nabla^2(\phi_{KS} k^\mu k_\nu)
		\Big],
		\label{ricci}
	\end{equation}
	which is manifestly linear in $\phi_{KS}$.
	
	The single-copy gauge field is defined by the identification \cite{Monteiro:2014cda}
	\begin{equation}
		A_\mu \equiv \phi_{KS}\, k_\mu.
		\label{Adef}
	\end{equation}
	Using this definition, Eq.~\eqref{ricci} can be rewritten as
	\begin{equation}
		{R^\mu}_\nu
		=
	{	\bar R^\mu}_{\ \, \nu}
		-\frac{1}{2}
		\left[
		\bar\nabla_\alpha F^{\alpha\mu}k_\nu
		+
		E^\mu_{\ \nu}
		\right],
		\label{riccif}
	\end{equation}
	where $F_{\mu\nu}=2\bar\nabla_{[\mu}A_{\nu]}$ and $E^\mu_{\ \nu}$ is the tensor
	\begin{align}
		{E^\mu}_\nu
		=&
		-\bar\nabla_\nu\!\left[
		A^\mu\!\left(
		\bar\nabla_\alpha k^\alpha
		+\frac{k^\alpha\bar\nabla_\alpha\phi_{KS}}{\phi_{KS}}
		\right)\right]
		+F^{\alpha\mu}\bar\nabla_\alpha k_\nu
		\nonumber\\
		&-\bar\nabla_\alpha
		\left(
		A^\alpha\bar\nabla^\mu k_\nu
		-A^\mu\bar\nabla^\alpha k_\nu
		\right)
		-\bar R^{\mu}_{\ \alpha\beta\nu}A^\alpha k^\beta
		+\bar R_{\alpha\nu}A^\alpha k^\mu.
		\label{etensor}
	\end{align}
	
	Substituting \eqref{riccif} into \eqref{reversed} and contracting with a Killing vector $V^\nu$ common to both the full and background metrics, one obtains the single-copy field equation
	\begin{equation}
		\bar\nabla_\nu F^{\nu\mu}
		+
		E^\mu
		=
		J^\mu,
		\label{single}
	\end{equation}
	where
	\begin{equation}
		E^\mu=\frac{1}{V\!\cdot\!k}\,E^\mu_{\ \nu}V^\nu,
	\end{equation}
	and
	\begin{equation}
		J^\mu
		=
		2\big(\Delta^\mu-\widetilde T^\mu\big),
		\qquad
		\Delta^\mu=\frac{1}{V\!\cdot\!k}\,\Delta^\mu_{\ \nu}V^\nu,
		\qquad
		\widetilde T^\mu=\frac{1}{V\!\cdot\!k}\,\widetilde T^\mu_{\ \nu}V^\nu.
	\end{equation}
	The ${\Delta^\mu}_ \nu$ tensor measuring the deviation from maximal symmetry is
	\begin{equation}
		{\Delta^\mu}_ \nu
		=
		\bar R^\mu_{\ \, \nu}
		-\frac{2\Lambda}{d-2}\,{\delta^\mu}_\nu.
		\label{delta}
	\end{equation}
	
	Contracting once more with $V^\mu$ yields the zeroth-copy equation,
	\begin{equation}
		\bar\nabla^2\phi_{KS}
		+
		\mathcal Z
		+
		\mathcal E
		=
		j,
		\label{zeroth}
	\end{equation}
	with
	\begin{equation}
		\mathcal Z=\frac{V\!\cdot\!C}{V\!\cdot\!k},
		\qquad
		\mathcal E=\frac{V\!\cdot\!E}{V\!\cdot\!k},
		\qquad
		j=\frac{V\!\cdot\!J}{V\!\cdot\!k},
	\end{equation}
	and
	\begin{equation}
		Z^\mu
		=
		\bar\nabla_\alpha k^\mu\,\bar\nabla^\alpha\phi_{KS}
		+
		\bar\nabla_\alpha
		\Big(
		2\phi_{KS}\,\bar\nabla^{[\alpha}k^{\mu]}
		-k^\alpha\bar\nabla^\mu\phi_{KS}
		\Big).
	\end{equation}
	
	For the static Lifshitz metrics introduced below, we take the time-like
	Killing vector $V^\mu=\delta^\mu_t$. With the Kerr--Schild vector $k^\mu$
	adapted to these metrics, one has $V\cdot k=1$, and $k^\mu$ is an
	eigenvector of the Killing two-form $\bar\nabla_\mu\bar V_\nu$, where
	$\bar V_\mu\equiv\bar g_{\mu\nu}V^\nu$. As shown in
	Appendix~\ref{app:Killing}, these two properties are sufficient to ensure
	that $E^{\mu}{}_{\nu}V^{\nu}=0$, without any use of the field equations.
	Consequently,
	\begin{equation}
		V\cdot k=1,\qquad E^\mu=0,\qquad \mathcal{E}=0.
	\end{equation}
	Thus, although the tensor $E^{\mu}{}_{\nu}$ need not vanish, $V^\nu$ is a
	right eigenvector of it with zero eigenvalue. As a result, the additional term drops out of both the single- and zeroth-copy equations. 
	
	We now specialize to static Lifshitz black holes,
	\begin{equation}
		ds^2 = 	\ell^2\left[ - r^{2z}f(r)\,dt^2	+\frac{dr^2}{r^2f(r)} +r^2 d\Sigma_{d-2,\kappa}^2 \right], \label{lifshitzline}
	\end{equation}
	with background metric obtained in the limit $f\to1$. After the coordinate transformation
	\begin{equation}
		dt\rightarrow dt+\frac{(f-1)r^{-(z+1)}}{f}\,dr,
	\end{equation}
	the metric assumes Kerr--Schild form with
	\begin{equation}
		\phi_{KS}=\ell^2 r^{2z} (1-f),	\qquad k_\mu \diff x^\mu = \diff t+\frac{\diff r}{r^{z+1}}. \label{phiandk}
	\end{equation}
	
	With this choice, the single- and zeroth-copy equations reduce to\footnote{For black holes, with planar horizon, the function of $\mathcal{Z}$ is decomposed in terms of Ricci scalar of background metric and KS scalar as:
		\begin{equation*}
		\mathcal{Z}=	\frac{(z-2)z}{z^2+2z+3}\,\bar{R}\,\phi_{KS}
		\end{equation*}
	 }
	\begin{align}
		\bar\nabla_\nu F^{\nu\mu}&=J^\mu, \label{Maxwell}
		\\
		\bar\nabla^2\phi_{KS}
		+
	\mathcal{Z}
		&=
		j.
	\end{align}
	
	This completes the Kerr--Schild construction for static Lifshitz black holes.
	
	Finally, we can conclude this discussion by writing the time component of Maxwell's equations in \eqref{Maxwell} for Lifshitz spacetime in explicit form. Let's start with gauge field defined in \eqref{Adef}:
	\begin{equation}
		A_\m = \f_{KS} k_\m\,.
	\end{equation}
	The time component of Maxwell equations can be written explicitly as
	\begin{align}
		\der_\m F^{\m t}&=J^t \nn \\
		-\frac{1}{\ell^4 r^{z+d-3}} \partial_r \left(r^{d-z-1} \f^\prime_{KS}(r)\right)&=\rho_{KS} \label{MaxwellT}
	\end{align}
	and this equation is the key point of the discussion in next section.
	
	\section{Minisuperspace Double Copy in Lifshitz Spacetimes} \label{MDC}

Before proceeding in this section, let us emphasize the conceptual difference with the Kerr--Schild (KS) approach. While the KSDC derives the single-copy structure from a specific geometric ansatz, our goal is instead to extract the same structure directly from the \emph{reduced action}, without assuming the existence of a Kerr--Schild decomposition. 
In this section, we construct the classical double copy within a one-dimensional minisuperspace reduction of Lifshitz spacetimes. Our goal is to show that the Maxwell operator can be extracted directly from the reduced gravitational action through a weighted radial operator, and that the corresponding charge density arises from the matter sector upon imposing the equations of motion. 
In this way, the minisuperspace construction provides an alternative route to the classical double copy, rooted in the constraint structure of symmetry-reduced gravity and applicable beyond maximally symmetric backgrounds.

In order to formulate general relativity in Hamiltonian language, one employs the Arnowitt--Deser--Misner (ADM) decomposition of the spacetime metric \cite{Arnowitt:1962hi}. 
In this framework, the line element is written as
\begin{equation}
	\diff{s}^2 = -N_\perp^2 \diff{t}^2 
	+ g_{ij}\left(\diff{x}^i + N^i \diff{t}\right)
	\left(\diff{x}^j + N^j \diff{t}\right),
\end{equation}
where $N_\perp$ denotes the lapse function, $N^i$ the shift vector, and $g_{ij}$ the induced metric on spatial hypersurfaces of constant time. With this decomposition, the Einstein--Hilbert action
\begin{equation}
	I_{\text{EH}} = \int \diff^d x \sqrt{-g}\, R
\end{equation}
can be expressed in Hamiltonian form as
\begin{equation}
	I_G = \int \diff^d x 
	\left[
	\Pi_{ij} \dot{g}^{ij}
	- N_\perp \mathcal{H}
	- N^i \mathcal{H}_i
	\right]
	+ B_G.
\end{equation}
Here, $\Pi_{ij}$ are the canonical momenta conjugate to $g^{ij}$, while $\mathcal{H}$ and $\mathcal{H}_i$ denote the Hamiltonian and diffeomorphism constraints, respectively. Owing to diffeomorphism invariance, the Hamiltonian contains no independent dynamical term; instead, the lapse and shift functions serve as Lagrange multipliers imposing the constraints $\mathcal{H}=0$ and $\mathcal{H}_i=0$. Further details of this construction can be found in the original ADM work and in subsequent reviews \cite{Banados:2016zim}.

Rather than pursuing a fully general treatment, we follow \cite{Crisostomo:2000bb} and restrict attention to static geometries with a maximally symmetric $(d-2)$-dimensional horizon of curvature $\kappa$. Such spacetimes can be derived from a one-dimensional minisuperspace reduction, and the relation between the integration constant and the associated conserved charge can be obtained using the Regge--Teitelboim approach \cite{Regge:1974zd}. 
However, in this work we do not focus on this issue. Instead, our interest is in how the reduced Hamiltonian structure organizes the double-copy construction.

To derive the reduced action, we substitute a static ansatz depending only on the radial coordinate
\begin{equation}
	ds^2 = \ell^2\left[- r^{2} f(r)\, N(r)^2\, dt^2 
	+ \frac{dr^2}{r^2 f(r)} 
	+ r^2 d\Sigma_{d-2,\kappa}^2 \right],
	\label{metricN}
\end{equation}
into the Einstein--Hilbert action. Here $d\Sigma^2_{d-2,\kappa}$ denotes the line element of a $(d-2)$-dimensional maximally symmetric space with curvature $\kappa$.

At this stage, the function $N(r)$ is kept arbitrary. It plays the role of a radial lapse and acts as a Lagrange multiplier enforcing the Hamiltonian constraint of the reduced system. Accordingly, the minisuperspace reduction is performed by first varying the action with respect to both $f(r)$ and $N(r)$, thereby deriving the full set of reduced equations, including the constraint.

Only after this step do we specialize the lapse. To implement Lifshitz asymptotics, we choose
\begin{equation}
	N(r)=r^{z-1}.
\end{equation}
With this ordering, the Lifshitz exponent $z$ enters through the choice of the lapse that solves the constraint for the matter content supporting the Lifshitz geometry. In particular, the relativistic limit $z\to1$ corresponds smoothly to the constant lapse $N(r)=1$.

After substituting the ansatz into the action and integrating by parts, the gravitational sector reduces to a total-derivative form,
\begin{equation}
	S^{g}_{red} =
	\ell^{d-2} (d-2)\Sigma_{d-2,\kappa}\Delta t
	\int \diff{r}\, N(r)\, \Psi^\prime(r)
	+ B_G,
	\label{redGrav}
\end{equation}
where we have introduced
\begin{equation}
	\Psi(r) =  \frac{r^{d-1-2z}}{\ell^2}\,\phi_{ms}(r).
\end{equation}

The quantity $\phi_{ms}(r)$, which we will refer to as the \emph{minisuperspace scalar}, is defined as
\begin{equation}
	\phi_{ms}(r)\equiv (\kappa - r^2)\,\ell^2 r^{2z-2} + \phi_{KS}(r),
	\label{ms31}
\end{equation}
where $\phi_{KS}$ is the Kerr--Schild scalar introduced in \eqref{phiandk}. This combination naturally emerges from the reduced action and packages both the Kerr--Schild contribution and the horizon-curvature term into a single effective potential. Unlike $\phi_{KS}$, which is tied to the Kerr--Schild decomposition, $\phi_{ms}$ is the natural variable for the minisuperspace formulation and therefore provides the appropriate potential for the minisuperspace double-copy construction.

In \eqref{redGrav}, we also introduce a boundary term $B_G$ which ensures a well-posed variational principle, such that the on-shell variation of the action vanishes. It is essential to include the function $N(r)$ in the metric ansatz; otherwise, the reduced action would consist purely of a boundary contribution and would not yield the equations determining the static solution. This minisuperspace reduction procedure applies to a broad class of gravitational theories, and several examples can be found in \cite{Deser:2003up}.
	
	Now we turn our attention to minisuperspace formalism of classical double copy and start with the action of the form
	\begin{equation}
		S=  \int \diff^dx  \sqdet \left(R - 2 \Lambda + \cL_m\right). \label{actionG}
	\end{equation}
	Here $\cL_m$ represents the matter Lagrangian density. To proceed, we switch to the one-dimensional reduced action and split it into two parts, gravitational and matter:
	\begin{equation}
		S_{red} = \ell^{d-2} 	\Sigma_{d-2,\kappa}\Delta t \int \diff r \left( \cLG+\cLM \right) .
	\end{equation}
	Here, $\cLG$ and $\cLM$ corresponds one-dimensional reduced Lagrangian densities of gravitational and matter parts of full theory, respectively, and we have also added the cosmological constant $\Lambda$ into the matter Lagrangian to compare our formalism with KS double copy easily. The reduced gravitational action 
	\begin{align}
		S_{red}^{\text {grav }}= &\ell^{d-2} \Sigma_{d-2, \kappa} \Delta t \int \diff r \cLG   \nn \\ 
		=&\ell^{d-2} \Sigma_{d-2, \kappa} \Delta t \int \diff r(d-2)\, N\, \partial_r\left(\frac{r^{d-1-2z}}{\ell^2}\phi_{ms}\right) \label{EactionKS}
	\end{align}
	
	As mentioned before, the reduced equations imply the Hamiltonian constraint, which fixes the on-shell relation between the reduced gravitational and matter sectors. If the matter Lagrangian is not linear in the lapse, the naive on-shell evaluation of $\cLG+\cLM$ can differ from zero by a total derivative related to boundary counterterms \cite{deHaro:2000vlm,Mann:2011hg}; the relevant object is $\cLN$ introduced below, and this point is illustrated in Sec.~\ref{LifshitzExamples}.
	
	At this point, in parallel with  \cite{Alkac:2024pfd}, the relation between $\cLG$ and the $tt$ component of the mixed-index Einstein equations can be expressed as follows: 
	\begin{equation}
		\cLG = -2\, \ell^2 r^{d-2} N(r) {G^t}_t. \label{GttLred}
	\end{equation}
	The relation in \eqref{GttLred} reduces to Eq.~(4.17) of \cite{Alkac:2024pfd} in the relativistic limit, obtained by setting $N(r)=r^{z-1}$, $z=1$, and $\ell=1$. Naturally, a similar relationship can be established with the energy-momentum tensor.
	
	In contrast to the gravitational sector, where $\cLG$ is linear in the lapse function $N(r)$, the reduced matter Lagrangian $\cLM$ is linear in $N(r)$ for most of its terms, but may also include non-linear powers of the lapse. Consequently, its variation with respect to $N(r)$ does not simply reduce to $\cLM/N(r)$. To put it more clearly, it would be incorrect to relate $\cLM/N(r)$ directly to the $tt$ component of the mixed-index energy-momentum tensor ${T^\mu}_{\nu}$. Nevertheless, the desired relation follows from the Hamiltonian constraint $\partial\cL_{red}/\partial N=0$. Since $\cLG=N\,h(r)$ is linear in $N$, on-shell this constraint reads $\cLG=-N\,\partial_N\cLM$, which coincides with $\cLG+\cLM=0$ whenever $\cLM$ is linear in $N$. We therefore define $\cLN\equiv N\,\partial_N\cLM$ (which equals $\cLM$ for terms linear in $N$) and write
	\begin{equation}
	\eval{\cLG+\cLN}_{\text{on-shell}}=0\,.
	\end{equation}

	We now proceed to formulate the discussion within the double-copy framework. To this end, we introduce the weighted radial operator
	\begin{equation}
		\mathcal{D}_z \equiv \frac{1}{b(r)}
		\left( r\partial_r + (2z-d+2) \right),
	\end{equation}
	which measures deviations from homogeneity of degree $d-2-2z$. Here the $b(r)$ is
	\begin{equation}
		b(r)= (d-2) \ell^2 r^{z+d-3} = (d-2) \frac{\sqdet}{\sqrt{\gamma}}\,,
	\end{equation}
	where $\sqrt{\gamma}$ is the square root of metric determinant which belongs to $d-2$ dimensional horizon. Both $b(r)$ and the weight in $\Dz$ are fixed by the Lifshitz lapse $N(r)=r^{z-1}$. In the definition of $\Dz$, the shift $2z-d+2$ incorporates the effective scaling weight determined by the Lifshitz exponent $z$ and the spacetime dimensionality $d$. In the case of $d=4$ and $z=1$, this operator reproduces Eq.~(5.4) of \cite{Alkac:2024pfd}.
	
The reduced gravitational sector for a static, radially dependent ansatz takes a universal divergence form
	\begin{equation}
		\cL^{g}_{\rm red}\;\propto\; N(r)\,\partial_r\!\Big(r^{d-1-2z}\,\mathcal{U}(r)\Big),
		\label{universal_divergence_form}
	\end{equation}
	where all theory dependence is packaged into a single ``minisuperspace potential'' $\mathcal{U}(r)$: $\mathcal{U}=\phi_{\rm ms}$ in Einstein gravity, and $\mathcal{U}=\mathcal{W}_{\rm ms}(\psi)$ in Lovelock theories \footnote{Lovelock's generalization of the formalism will be examined in detail later in this section.}. After specializing to the Lifshitz lapse $N=r^{z-1}$, the natural notion of radial homogeneity is controlled by the weight $(2z-d+2)$. The combination $r\partial_r+(2z-d+2)$ therefore measures the failure of $\mathcal{U}(r)$ to scale with the Lifshitz weight implied by the ansatz, while the normalization by $b(r)\propto \sqdet/\sqrt{\gamma}$ is chosen so that the resulting operator matches precisely the canonical Maxwell form $\ell^{-4} r^{-(z+d-3)}\partial_r(r^{d-z-1}\partial_r\,\cdot\,)$ acting on the single-copy potential. In this way $\mathcal{D}_z$ is not engineered for a particular solution but is fixed by the universal reduced-action structure and by Lifshitz scaling weights.
	
	Acting with this operator on the reduced gravitational Lagrangian, we obtain the identity
	\begin{align}
		\mathcal{D}_z \cLG + \mathcal{Z}_{\mathrm{ms}}(r)=& \frac{1}{\ell^4 r^{z+d-3}} \partial_r \left( r^{d-z-1} \bar\phi_{ms}^\prime(r) \right)\nn \\
		=&-\bar{\der}_\m F^{\m t}
		\label{msMaxwell}
	\end{align}
	where the function $\mathcal{Z}_{\mathrm{ms}}(r)$ collects the residual contributions arising from the non-maximally symmetric Lifshitz background within the minisuperspace reduction. In \eqref{msMaxwell} the $F_{\m\n}$ is the field strength tensor of minisuperspace single copy and it has the form of
	\begin{equation}
		F_{\m\n} =\partial_\m A^{(ms)}_\n - \partial_\n A^{(ms)}_\m \qquad\text{where}\qquad A^{(ms)}_\m \diff x^\m = \bar\phi_{ms} \diff t. 
	\end{equation}  
	
	The right-hand side in \eqref{msMaxwell} reproduces the same radial operator that appears in the temporal component of Maxwell’s equations in \eqref{MaxwellT}, although the latter is written in terms of the Kerr--Schild scalar. Hence, up to the additional term $\mathcal{Z}_{\mathrm{ms}}(r)$, the Maxwell operator emerges directly from the gravitational minisuperspace Lagrangian. Here the single-copy potential $\bar\phi_{ms}\equiv\phi_{ms}-\kappa\ell^2r^{2z-2}=\phi_{KS}-\ell^2r^{2z}$ differs from $\phi_{ms}$ only by the horizon-curvature piece of Eq.~\eqref{ms31}; the Maxwell-operator image of this piece is exactly the $\kappa$-dependent term of $\mathcal{Z}_{\mathrm{ms}}$ in Eq.~\eqref{Zms}. For $z=1$ the two potentials differ by a constant and have the same field strength.	This relation holds off shell in the sense that no use has been made of the field equations or of their explicit solutions. However, it is not completely independent of the underlying dynamics. It relies on the universal structure of the reduced Lagrangian, whose radial dependence is governed by the lapse function multiplying a total-derivative term. The quantity inside this derivative is directly related to the radial Hamiltonian constraint and, equivalently, to the $tt$ component of the mixed-index Einstein equations. Thus, the identity should be understood as a kinematical statement within the symmetry-reduced framework, obtained after specializing the lapse and integrating by parts, rather than as a consequence of solving the full equations of motion.
	
	We now analyze the residual term $\mathcal{Z}_{\mathrm{ms}}(r)$. Its explicit form is
	\begin{equation}
		\mathcal{Z}_{\mathrm{ms}}(r)
		=
		-\frac{(z-1)(d-2z-1)}{\ell^4}\,
		\frac{\phi_{ms}(r)}{r^{2z}}  -\frac{2 \kappa (z-1)(d+z-4)}{\ell^2 r^2}.
		\label{Zms}
	\end{equation}
	Several structural features follow directly from \eqref{Zms}. The term proportional to $\phi_{ms}(r)$ carries the universal prefactor $(z-1)(d-2z-1)$ and therefore vanishes in the relativistic limit $z=1$. In this sense, $\mathcal{Z}_{\mathrm{ms}}(r)$ measures the deviation from AdS induced by anisotropic Lifshitz scaling. Its form is fixed by radial weight counting and diffeomorphism invariance: the factor $\phi_{\mathrm{ms}}/r^{2z}$ is required by dimensional consistency, while $(d-2z-1)$ encodes the homogeneity mismatch of the reduced potential.
	
	The second term, proportional to $\kappa$, equals minus the radial Maxwell operator acting on the horizon-curvature piece $\kappa\ell^2 r^{2z-2}$ of $\phi_{ms}$. Its role is to remove this background piece from the single-copy potential, so that $\kappa$ enters the charge density only through $\phi_{KS}$. It is proportional to $(z-1)$, so it disappears at $z=1$, where $\kappa\ell^2$ is a constant that does not contribute to the field strength anyway.
	
	Accordingly, $\mathcal{Z}_{\mathrm{ms}}$ is not an algebraic remainder introduced to force a Maxwell-like form. The same prefactor structure reappears in Lovelock theories, indicating that it is determined by scaling and radial diffeomorphism invariance rather than by theory-specific tuning. This is consistent with the AdS limit: when $z\to1$, Lifshitz reduces to maximally symmetric AdS, the residual term vanishes, and the reduced identity collapses to the pure Maxwell operator. The $\kappa$-dependent term is not an independent ingredient: it is fixed by the horizon-curvature piece of $\phi_{ms}$.
	
	We now explain why we define a \emph{minisuperspace scalar} $\phi_{ms}$. Using $\phi_{ms}$, rather than only the Kerr--Schild scalar $\phi_{KS}$, is not merely a matter of notation. First, $\phi_{ms}$ is the variable in which the reduced Lagrangian takes the divergence form \eqref{EactionKS} for all $\kappa$; for $\kappa\neq0$ it contains the horizon-curvature piece $\kappa\ell^2r^{2z-2}$, which is not part of the Kerr--Schild sector. If one worked only with $\phi_{KS}$, the identity could not be read off directly from the reduced Lagrangian. Second, $\phi_{ms}$ provides a more natural comparison with the KSDC, as it combines the geometric contribution from $(\kappa - r^2)$ with the Kerr--Schild sector into a single effective potential adapted to the reduced-action framework. Finally, this definition is essential for higher-curvature generalizations, where the role of $\phi_{ms}$ is played by the minisuperspace Wheeler polynomial, ensuring that both the geometric contribution and the correct radial scaling are preserved.
	
	Our next task is to understand how the matter sector enters the construction. First, we note that $\Dz$ is a linear operator. Second, the total reduced Lagrangian satisfies
	\begin{equation}
		\eval{\cLG+\cLN }_{\text{on-shell}} = 0.
	\end{equation}
	By linearity of $\Dz$, this implies
	\begin{align}
		\Dz \cLG &=\eval{ - \Dz \cLN}_{\text{on-shell}}  \nn \\
		\Dz \cLG +\Zms  &=\eval{ - \Dz \cLN +\Zms}_{\text{on-shell}} 
 	\end{align}
 	We emphasize that the identity
 \begin{equation}
 		\Dz \cLG + \Zms = -\der_\mu F^{\mu t}
 \end{equation}
 	holds off-shell and follows purely from the structure of the minisuperspace reduction. However, substituting the on-shell relation between $\cLG$ and $\cLM$ into this identity yields
 	\begin{equation}
 		\nabla_\mu F^{\mu t} = \eval{\Dz \cLN -\Zms }_{\text{on-shell}}
 		\label{DCdenkleme}
 	\end{equation}
 	Thus, while the left-hand side is defined off-shell as the Maxwell operator, its interpretation as a sourced Maxwell equation arises only after imposing the gravitational equations of motion. Consistency then requires that the right-hand side be identified with the charge density of the Kerr–Schild single copy,
	\begin{equation}
		\Dz \cLN -\Zms= \eval{\rho_{ms} }_{\text{on-shell}} 
	\end{equation}
	where
	\begin{equation}
	J^t=\rho_{ms}(r),\qquad \rho_{ms}(r)=\rho_{KS}(r)+\Delta_{ms} \qquad\text{and}\qquad \Delta_{ms}=\frac{2z(z+d-2)}{\ell^2}\,.
	\end{equation}
	Here the $\rho_{KS}(r)$ is the charge distribution of KS zeroth copy. In addition, the $\Delta_{ms}$ gives a constant charge density which is equivalent to KS counterpart ${\Delta^\m}_\n$ which is defined in \eqref{delta}. However in our formalism we directly assume that the cosmological constant as a part of matter Lagrangian density. This is a bookkeeping choice: $ \Lambda$ is moved to $\cLM$ so that $\Delta_{ms}$ measures only the non-maximally-symmetric Lifshitz deviation beyond the cosmological constant.
	
	With this identification, the minisuperspace double-copy construction is complete. The construction mirrors the KSDC at the level of the reduced action, with the minisuperspace scalar $\phi_{ms}$ playing the role of the effective single-copy potential. The Maxwell operator emerges off-shell from the gravitational minisuperspace Lagrangian through the weighted radial operator $\Dz$, while the corresponding charge density arises on-shell from the matter sector together with the residual Lifshitz contribution $ \Zms$. In this way, the double-copy correspondence is realized at the level of the reduced action, with deviations from maximal symmetry encoded explicitly in the Lifshitz-dependent term $\Zms$.
	
	Our minisuperspace construction applies to static ans\"atze that depend only on the radial coordinate and have a maximally symmetric $(d-2)$-dimensional horizon, for which the reduced action takes a universal divergence form dictated by radial diffeomorphism invariance. In this sector, the Maxwell operator emerges off shell from the reduced gravitational Lagrangian, while the source is fixed on shell by the Hamiltonian constraint. The KSDC, by contrast, requires the existence of a KS decomposition with a null geodesic vector and a suitable Killing vector common to both the background and full metrics. For static Lifshitz black holes admitting such KS coordinates (with $V^\mu=\partial_t$), the two constructions overlap, and the on-shell source extracted from minisuperspace reproduces the corresponding KSDC single-copy charge. Outside this overlap, neither framework is expected to be universally applicable: there exist KS spacetimes that are not captured by our symmetry-reduced ansatz, and conversely, there are symmetry-reduced configurations for which a KS form is not available globally.

We now extend the construction to Lovelock theories, whose field equations remain second order in derivatives of the metric \cite{Lovelock:1971yv}. This provides a non-trivial consistency check and demonstrates that the minisuperspace double-copy structure is not restricted to two-derivative gravity. The total action of a general $n$th-order Lovelock theory is
\begin{equation}
	S^{LL} = \int \diff^d x \sqdet   \sum^{\left[\frac{d-1}{2}\right]}_{n=1} \alpha_n\cL_n . \label{lovelockAction}
\end{equation}
here $\cL_n$ is the $n$'th order Lovelock Lagrangian density in the form
\begin{equation}
	\mathcal{L}_n= \frac{1}{2^n} \delta^{a_1 b_1 a_2 b_2 \dots a_n b_n}_{c_1 d_1 c_2 d_2 \dots c_n d_n} \prod_{p=1}^{n} R^{c_p d_p}{}_{a_p b_p},
\end{equation}
and $\alpha_n$'s are the coupling constants of n'th order Lovelock Lagrangian.

For this class of theories, the reduced action takes a simple total-derivative form governed by the Wheeler polynomial  \cite{Wheeler:1985nh,Wheeler:1986zp}. The action in \eqref{lovelockAction} can be written for the metric ansatz in \eqref{lifshitzline}
	\begin{equation}
		S^{LL}_{red} =
		\ell^{d-2} (d-2)\Sigma_{d-2,\kappa}\Delta t
		\int \diff{r}\, N\, \partial_r \left(r^{d-1} \mathcal{W}(\psi) \right)
		+ B_G,
	\end{equation}
where $\mathcal{W}(\psi)$ is Wheeler polynomial in the form
	\begin{equation}
		\mathcal{W}(\psi) =  \sum_{n=1} \alpha_n \psi^n\qquad\text{and}\qquad \psi \equiv \frac{\kappa }{r^2} - f(r)\label{wheelerP}
	\end{equation}
 To write the reduced action in double copy language, the Wheeler polynomial should be modified by using the definition of $\phi_{KS}$ in \eqref{phiandk}. In this case, the action reads
 	\begin{equation}
 	S^{LL}_{red} =
 	\ell^{d-2} (d-2)\Sigma_{d-2,\kappa}\Delta t
 	\int \diff{r}\, N\, \Psi^\prime_{LL}
 	+ B_G, \label{LLactionKS}
 \end{equation}
 where
 \begin{equation}
 	 \Psi_{LL} =  \frac{1}{\ell^2}r^{d-1-2z} \mathcal{W}_{ms}(\psi) 
 \end{equation}
and we  can define a new quantity, the minisuperspace Wheeler polynomial as
\begin{equation}
	\mathcal{W}_{ms}(\psi) = \ell^2 r^{ 2z} \sum_{n=1} \hat{\alpha}_n \psi^n \qquad\text{and}\qquad \psi = \frac{\kappa-r^2}{r^2} +\frac{\f_{KS}}{\ell^2 r^{2z}}\,.
	 \label{wheelerPms}
\end{equation}
Here, the $\hat{\alpha}_n$ are the appropriately dimensionless versions of the couplings $\alpha_m$. The minisuperspace Wheeler polynomial $\mathcal{W}_{ms}(\psi)$ defined here will play a pivotal role in our minisuperspace double-copy formalism.

We can now transparently compare the reduced Einstein–Hilbert action in \eqref{EactionKS} with the Lovelock reduced action in \eqref{LLactionKS} to observe how higher-curvature terms modify the single-copy potential. First, it is immediately apparent that the newly defined weighted Wheeler polynomial in \eqref{LLactionKS} acts as an effective minisuperspace potential, analogous to \eqref{EactionKS}. Second, we notice that the $\kappa-r^2$ term is now entirely absorbed into $\mathcal{W}_{ms}$. This demonstrates that a polynomial of the $\kappa$ parameter contributes as a charge in the double copy formalism. This indicates that higher-curvature terms modify the effective minisuperspace potential and hence the asymptotic charge bookkeeping, while the off-shell Maxwell-operator identity remains purely kinematical. Finally, the residual term for $\mathcal{W}_{ms}$, $ \mathcal{Z}^{LL}_{ms}$ takes a simple form:
\begin{equation}
	\mathcal{Z}^{LL}_{ms}= -\frac{(z-1)(d-2z-1)}{\ell^4} \frac{\mathcal{W}_{ms}}{r^{2z}}  -\frac{2 \kappa (z-1)(d+z-4)}{\ell^2 r^2},
	\label{ZmsLL}
\end{equation}
which matches \eqref{Zms} upon the replacement $\phi_{ms}\rightarrow \mathcal{W}_{ms}$, with the same additional horizon-curvature term proportional to $\kappa$.

The persistence of the residual structure in Lovelock theories is not a trivial consequence of polynomiality. Although the Wheeler polynomial modifies the functional form of the reduced potential and encodes higher-curvature contributions, the reduced action retains the universal divergence structure dictated by radial diffeomorphism invariance. The weighted operator $\Dz$ acts on this universal constraint form rather than on the specific polynomial details. The residual term decomposes into a potential-dependent contribution with the universal prefactor $(z-1)(d-2z-1)$, and a piece proportional to $\kappa$ that equals minus the Maxwell operator acting on $\kappa\ell^2r^{2z-2}$. Both contributions are fixed by the radial scaling weights in Lifshitz and by radial diffeomorphism invariance of the reduced system, rather than by the detailed polynomial structure of $\mathcal{W}(\psi)$.

A potential source of confusion is the following: in Lovelock theories the effective minisuperspace potential is the weighted Wheeler polynomial
$\mathcal{W}_{\rm ms}(\psi)$, and since $\psi$ contains the combination $\kappa/r^2$, $\mathcal{W}_{\rm ms}$ generically involves \emph{polynomials} in $\kappa$.
One might then expect the residual term $\mathcal{Z}_{\rm ms}$ to inherit higher powers of $\kappa$ as well. 
This is \emph{not} the case, because the $\kappa$-dependent term of $\mathcal{Z}_{\rm ms}$ is the Maxwell-operator image of the single piece $\kappa\ell^2r^{2z-2}$, which does not depend on the functional form of the potential.

More precisely, the off-shell identity
\begin{equation}
	\mathcal{D}_z \cL^{g}_{\rm red}+\mathcal{Z}_{\rm ms}(r)=\frac{1}{\ell^4 r^{z+d-3}}\,\partial_r\!\Big(r^{d-z-1}\,\bar{\mathcal{U}}'(r)\Big)=-\nabla_\mu F^{\mu t},
	\qquad
	\bar{\mathcal{U}}\equiv\mathcal{U}-\kappa\ell^2r^{2z-2},
\end{equation}
where $F_{\mu\nu}$ is built from $A_t=\bar{\mathcal{U}}$, is derived by acting with the weighted operator $\mathcal{D}_z$ on the \emph{universal divergence form} of the reduced action,
\begin{equation}
	\cL^{g}_{\rm red} \;\propto\; N(r)\,\partial_r \Big(r^{d-1-2z}\, \mathcal{U}(r)\Big),
	\qquad 
	\mathcal{U}(r)=
	\begin{cases}
		\phi_{\rm ms}(r) & \text{(Einstein)}\\[2pt]
		\mathcal{W}_{\rm ms}(\psi(r)) & \text{(Lovelock)}~,
	\end{cases}
\end{equation}
and uses only integration by parts together with radial scaling weights fixed by the Lifshitz ansatz $N(r)=r^{z-1}$.
In this manipulation, all theory-dependent information is encoded in the single object $\mathcal{U}(r)$, producing the potential-dependent piece in $\Zms$
\begin{equation}
  	-\frac{(z-1)(d-2z-1)}{\ell^4}\,\frac{\mathcal{U}(r)}{r^{2z}}\,,
\end{equation}
which indeed inherits whatever $\kappa$-polynomial structure is present in $\mathcal{U}$ (e.g.\ in $\mathcal{W}_{\rm ms}$).

By contrast, the term proportional to $\kappa$ does not depend on $\mathcal{U}$. Writing $\mathcal{U}=\bar{\mathcal{U}}+\kappa\ell^2r^{2z-2}$, the Maxwell operator acting on the second piece gives $2\kappa(z-1)(d+z-4)/(\ell^2r^2)$, and the corresponding term in $\Zms$ removes it,
\begin{equation}
	\mathcal{Z}_{\rm ms}(r)\supset -\frac{2\kappa (z-1)(d+z-4)}{\ell^2 r^2}\,.
\end{equation}
Higher powers such as $\kappa^2/r^4$ are not produced by this term: they belong to $\mathcal{U}(r)$ and, if present, are already contained in $\mathcal{W}_{\rm ms}(\psi)$.
Therefore, even in Gauss--Bonnet or more general Lovelock theories, the residual term decomposes universally into
(i) a potential-dependent part that may carry $\kappa$-polynomials via $\mathcal{U}$, and (ii) a piece that is \emph{linear} in $\kappa$, scales as $r^{-2}$, and removes the horizon-curvature piece $\kappa\ell^2r^{2z-2}$ from the single-copy potential.
Both pieces vanish at $z=1$ and for $\kappa\to0$ the second one is absent, as required by the relativistic and planar limits.


As a summary for this section, we have established a robust minisuperspace double-copy framework for static Lifshitz spacetimes with maximally symmetric horizons. By employing the weighted radial operator $\Dz$, we demonstrated that the Maxwell operator emerges directly from the reduced Einstein-Hilbert action off-shell, while the corresponding charge density is recovered on-shell from the matter sector. A crucial feature of this construction is the residual term $\Zms$, which explicitly captures the deviation from exact maximal symmetry introduced by the anisotropic scaling $z\neq1$. Furthermore, our generalization to higher-order Lovelock gravity shows that the same structure appears in this procedure. The modified Wheeler polynomial $\mathcal{W}_{ms}$ naturally assumes the role of the minisuperspace potential $\phi_{ms}$, while preserving the compact structure of the residual deviation $\mathcal{Z}^{LL}_{ms}$. This indicates that the minisuperspace double-copy correspondence extends naturally to higher-curvature theories even in non-relativistic backgrounds. Hence the minisuperspace double copy is not an accident of two-derivative gravity, but rather a structural property of the radial Hamiltonian constraint itself.


\section{Static spherical sector as a reference configuration} \label{static_spherical}

In this section, we consider the static spherically symmetric vacuum sector as a reference configuration for the minisuperspace construction. 
In contrast to the Lifshitz case, no anisotropic scaling is present ($z=1$), and the geometry is fully characterized by the spherical horizon curvature. 
This provides a clean setting in which the reduced action and its relation to the Maxwell operator can be made completely explicit.

We set $\kappa=1$ in \eqref{metricN} and fix $\ell=1$, obtaining
\begin{equation}
	ds^2 = - r^{2} f(r)\, N(r)^2\, dt^2 
	+ \frac{dr^2}{r^2 f(r)} 
	+ r^2 d\Omega_{d-2}^2.
\end{equation}
As in the general construction, the function $N(r)$ is kept arbitrary during the variation and enforces the Hamiltonian constraint.

Substituting the ansatz into the Einstein--Hilbert action and integrating by parts, the reduced gravitational action takes the universal form
\begin{equation}
	S^{g}_{red} = \int dr\, N(r)\, \Psi'(r) + B_G,
\end{equation}
with
\begin{equation}
	\Psi(r) = r^{d-3}\,\phi_{ms}(r).
\end{equation}

In this sector, the minisuperspace scalar decomposes as
\begin{equation}
	\phi_{ms}(r) = (1-r^2) + \phi_{KS}(r),
\end{equation}
where $(1-r^2)$ encodes the background curvature contribution, while $\phi_{KS}$ is defined as
\begin{equation}
	\phi_{KS}(r)=r^2\bigl(1-f(r)\bigr).
\end{equation}

The single-copy gauge potential is defined directly as
\begin{equation}
	A_t(r)=\phi_{ms}(r),
\end{equation}
Since $\bar\phi_{ms}=\phi_{ms}-\kappa$ differs from $\phi_{ms}$ only by a constant at $z=1$, the associated Maxwell operator takes the form
\begin{equation}
	\nabla_\mu F^{\mu t}
		=
		-\frac{1}{r^{d-2}}\partial_r\!\left(r^{d-2}\partial_r \phi_{ms}(r)\right).
\end{equation}
Acting with the reduced operator $\mathcal D_1$ on the gravitational Lagrangian then yields the off-shell identity
\begin{equation}
	\mathcal D_1 \cLG
		=
		-\nabla_\mu F^{\mu t},
\end{equation}
showing that the Maxwell operator emerges directly from the reduced gravitational sector.

For the vacuum Schwarzschild solution in this parametrization,
\begin{equation}
	r^2 f(r)=1-\frac{2\mu}{r^{d-3}},
\end{equation}
or equivalently
\begin{equation}
	f(r)=\frac{1}{r^2}-\frac{2\mu}{r^{d-1}}.
\end{equation}
This implies
\begin{equation}
	\phi_{KS}(r)=r^2(1-f)=r^2-1+\frac{2\mu}{r^{d-3}},
\end{equation}
and therefore
\begin{equation}
	\phi_{ms}(r)=\frac{2\mu}{r^{d-3}}.
\end{equation}

Imposing the Hamiltonian constraint $\Psi'(r)=0$, one obtains
\begin{equation}
	\nabla_\mu F^{\mu t}=0,
\end{equation}
so that the effective charge density vanishes in the bulk,
\begin{equation}
	\rho_{ms}(r)=0 \qquad (r\neq 0),
\end{equation}
as expected for a vacuum solution. 
The parameter $\mu$ is thus associated with a localized source at the origin, which is not resolved within the minisuperspace reduction.

The relation between the minisuperspace and Kerr--Schild charge densities,
\begin{equation}
	\rho_{ms}(r)=\rho_{KS}(r)+\Delta_{ms},
\end{equation}
provides a useful consistency check. 
For $z=1$ and $\ell=1$, one has
\begin{equation}
	\Delta_{ms}=2(d-1),
\end{equation}
while evaluating the background Maxwell operator on $\phi_{KS}$ yields the Kerr--Schild charge density
\begin{equation}
	\rho_{KS}=-2(d-1).
\end{equation}
Here $\Delta_{ms}$ compensates the contribution of the reference term $(1-r^2)$ in $\phi_{ms}$. These contributions cancel exactly, giving
\begin{equation}
	\rho_{ms}=0,
\end{equation}
in agreement with the vacuum minisuperspace equation.

It is worth noting that the expressions obtained here may look different from the standard forms of the Schwarzschild solution and its associated single-copy quantities commonly used in the literature (see e.g.\ \cite{Alkac:2021bav,CarrilloGonzalez:2017iyj}). This difference is purely a consequence of the parametrization adopted in \eqref{metricN}, where the metric functions appear in the combinations $r^2 f(r)$ and $1/(r^2 f(r))$. Compared to the more conventional Schwarzschild-like coordinates, this introduces additional $r^2$ factors into the definitions of $\phi_{KS}$ and $\phi_{ms}$, effectively redistributing constant and radial contributions between background and dynamical pieces.

When rewritten in the usual parametrization, these expressions map directly onto the standard results reported in \cite{Alkac:2021bav,CarrilloGonzalez:2017iyj}. In particular, the vanishing of the bulk charge density $\rho_{ms}=0$ is fully consistent with the vacuum nature of the solution. The apparent differences are therefore purely representational and do not reflect any physical mismatch.

This demonstrates that, in the static spherical sector, the minisuperspace scalar $\phi_{ms}$ absorbs the constant background contribution present in the Kerr--Schild description, leaving a sourceless Maxwell equation in the bulk. 
The Lifshitz case considered below can then be understood as a deformation of this structure induced by anisotropic scaling.

	\section{Lifshitz Black Hole Examples}\label{LifshitzExamples}
	
	To illustrate the structural content of the minisuperspace double copy, we now consider three representative classes of Lifshitz black hole solutions. These examples are chosen to probe complementary aspects of the formalism: the role of matter sourcing, the emergence of charge densities, and the effect of horizon topology on the reduced radial dynamics.
	
	The first example involves a Lifshitz solution supported by a scalar field conformally coupled to a Maxwell sector. In the Kerr--Schild construction, this configuration leads to a vacuum single copy\cite{Alkac:2021bav}. It therefore provides a non-trivial test of whether the minisuperspace operator reproduces a vanishing Maxwell source off-shell, despite the presence of matter in the gravitational description.
	
	The second example consists of a charged Lifshitz black hole supported by a Maxwell and a Proca field. In this case, the KS single copy produces a non-vanishing charge density associated with the Maxwell sector \cite{Alkac:2021bav}. Within our formulation, this solution allows us to verify that the Hamiltonian constraint dynamically generates the expected sourced Maxwell equation, thereby realizing the on-shell completion of the off-shell geometric identity derived in the previous section.
	
	Finally, we turn to a topological Lifshitz black hole. Unlike the previous examples, the horizon geometry is not restricted to the planar case, allowing us to examine how non-trivial horizon topology modifies the reduced Wheeler polynomial and the residual term appearing in the minisuperspace double copy. In this setup, we further introduce an additional electric charge, enabling a simultaneous analysis of topological contributions and matter-induced sourcing. This example is particularly relevant from a holographic perspective, as the interplay between anisotropic scaling, horizon curvature, and charge affects the structure of boundary counterterms in holographic renormalization.
	
	Taken together, these three examples demonstrate that the minisuperspace double copy is sensitive not only to matter content but also to geometric data such as horizon topology, while preserving the universal structure of the radial Maxwell operator.
	
\subsection{Lifshitz Black Hole sourced from a Massless Scalar and a Gauge Field }	
	 Our first example stems from the matter Lagrangian \cite{Taylor:2008tg}:
	\begin{equation}
		\cL_m=\frac{1}{2} \partial_\mu \varphi \partial^\mu \varphi-\frac{1}{4} e^{\lambda \varphi} f_{\mu \nu} f^{\mu \nu}.
	\end{equation}
	and matter configuration can be given as
	\begin{equation}
		\begin{aligned}
			f_{r t} & =q e^{-\lambda \varphi} r^{z-d+1}, \quad e^{\lambda \varphi}=r^{\lambda \sqrt{2(z-1)(d-2)}}, \quad
			\lambda^2 =\frac{2(d-2)}{z-1},\\
			 q^2&=2 \ell^2(z-1)(z+d-2), \quad \Lambda  =-\frac{(z+d-3)(z+d-2)}{2 \ell^2} .
		\end{aligned}
	\end{equation}
	The metric function $f(r)$ is in the form:
	\begin{equation}
		f(r)=1-\left(\frac{r_+}{r}\right)^{d+z-2}\,.
	\end{equation} 
	For this solution the $\phi_{KS}$ and $\phi_{ms}$ can be calculated as
	\begin{equation}
		\phi_{KS}(r)= \ell^2 r^{2z} \left(1 -f(r)\right)=  \ell^2\,\frac{r^{d+z-2}_+}{r^{d-z-2}} \qquad\text{and} \qquad \phi_{ms}(r) =-\ell^2 r^{2z}+\f_{KS}(r) 
	\end{equation}
	and the potential $\phi_{KS}$ solves the vacuum Maxwell equation, whereas $\phi_{ms}$ is sourced by the constant charge density below. Namely, the zeroth-copy charge densities of KS and minisuperspace formalism are
	\begin{equation}
	\rho_{KS} = 0\qquad \text{and} \qquad \rho_{ms}(r) = \frac{2z(z+d-2)}{\ell^2}.
	\end{equation}
	This result is consistent with the existing double-copy literature \cite{Alkac:2021bav}.
	
	For $z\to1$ the metric function reduces to $f(r)=1-(r_+/r)^{d-1}$, which is precisely the AdS--Schwarzschild solution. In this limit the deviation term $\mathcal{Z}_{\rm ms}$ vanishes identically, and the minisuperspace identity collapses to the pure Maxwell operator. The residual constant contribution in $\rho_{\rm ms}$ corresponds to the background curvature (cosmological constant) term, which in the Kerr--Schild formulation is encoded in ${\Delta^\mu}_{\nu}$. Hence the $z\to1$ limit is fully consistent with the standard classical double copy in AdS.
	
	\subsection{Lifshitz Black Hole sourced from a Proca and a Gauge Field }
	
	The matter Lagrange density of the second example is 
	\begin{equation}
		\cL_m=-\frac{1}{4} f_{\mu \nu} f^{\mu \nu}-\frac{1}{2} m^2 a_\mu a^\mu-\frac{1}{4} \mathcal{F}_{\mu \nu} \mathcal{F}^{\mu \nu}.
	\end{equation}
where $a_\m$ is a massive vector field with the field strength $f_{\m\n} = 2 \partial_{\left[\m\right.} a_{\left.\n\right]}$ and $\mathcal{F}_{\mu \nu}$ is the field strength of the gauge field \cite{Pang:2009pd}. The matter configuration
\begin{equation}
	a_t=\ell \sqrt{\frac{2(z-1)}{z}} f(r) r^z, \quad \quad \mathcal{F}_{r t}=q \ell r^{z-d-1}
\end{equation}
To satisfy field equations, one should fix the mass of the Proca field, the cosmological constant and the Lifshitz exponent as follows
\begin{equation}
	m=\sqrt{\frac{(d-2) z}{\ell^2}}, \qquad \Lambda=-\frac{(d-3) z+(d-2)^2+z^2}{2 \ell^2}, \qquad 	z= 2\left(d-2\right).
\end{equation}
These matter fields yield a metric function in the form
\begin{equation}
	f(r)=1-\frac{q^2}{2 (d-2)^2 r^z}. \label{met2}
\end{equation}
To proceed with the double-copy analysis, we first obtain the KS and minisuperspace scalars. We begin with the KS formalism.
\begin{align}
	\phi_{KS}(r) =& \ell^2 r^{2z} \left(1-f(r)\right)\nn \\
	 =&\frac{\ell^2 q^2 r^{z}}{2 \left(d-2\right)^2}.
\end{align}
and this zeroth copy potential gives a single-copy electric field
\begin{equation}
	F_{rt} = \frac{\ell^2 q^2 r^{z-1}}{2(d-2)^2}
\end{equation}
and it satisfies the Maxwell equation
\begin{equation}
	\nabla_\nu F_{KS}^{\nu \mu}=J_{KS}^\mu, \qquad \text{with} \qquad 	J_{KS}^\mu =  -\frac{q^2}{ \ell^2 r^z} {\delta^\mu}_0.
\end{equation}
We now turn to the minisuperspace double-copy construction. The minisuperspace potential is
\begin{align}
	\phi_{ms} =& - \ell^2 r^{2z} +\f_{KS} \nn \\
	=&- \ell^2 r^{2z}+\frac{\ell^2 q^2 r^{z}}{2 \left(d-2\right)^2}\,.
\end{align}
This scalar potential is sourced from a charge density
\begin{equation}
	\rho_{ms}(r)= -\frac{q^2}{ \ell^2 r^z}+\frac{2z(z+d-2)}{\ell^2}.
\end{equation}

Unlike the first example, this solution does not admit a smooth relativistic limit. The Lifshitz exponent is fixed by the field equations to $z=2(d-2)$, and therefore $z\to1$ cannot be achieved without leaving the solution space. Consequently the geometry does not reduce to an AdS black hole, and the minisuperspace charge density retains its intrinsic Lifshitz character.

\subsection{Lifshitz Topological Black Hole}
As a final example, we consider the charged Lifshitz topological black hole, introduced in \cite{Mann:2009yx} and further analyzed in \cite{Brynjolfsson:2009ct}.

The horizon topology plays a central role in this solution. In $d=4$, the horizon metric is
	\begin{equation}
	\dif \Sigma^2_{d=4,\kappa} = \begin{cases}
		\dif \theta^2 + \text{sin}^2(\theta) \dif \vf^2, &\kappa =1 \\
		\dif \theta^2 + \theta^2 \dif \vf^2, &\kappa=0 \\
		\dif \theta^2 + \text{sinh}^2(\theta) \dif \vf^2, &\kappa =-1 \\
	\end{cases} .
\end{equation}

The matter Lagrangian is
\begin{equation}
	\cL_m = - \frac{1}{4}\mathcal{F}_{\m\n} \mathcal{F}^{\m\n} - \frac{1}{12} \cH_{\m\n\rho} \cH^{\m\n\rho} - C \epsilon^{\m\n\rho\sigma} \mathcal{B}_{\m\n}\mathcal{F}_{\rho\sigma} - \frac{1}{4} \mathcal{G}_{\m\n} \mathcal{G}^{\m\n}.\label{topLm}
\end{equation}
Here $\mathcal{F}_{\mu\nu}$ is a two-form and $\mathcal{H}_{\mu\nu\sigma}=3\partial_{[\mu}\mathcal{B}_{\nu\sigma]}$ is a three-form; they are topologically coupled through the constant $C$. The additional two-form $\mathcal{G}_{\mu\nu}$ generates the charged branch studied here.

The solution is supported by the matter configuration
\begin{equation}
	\mathcal{F}_{rt} = - \frac{\sqrt{6} \ell }{10}  \left(20 r^3+\kappa r\right),\qquad \cH_{r\theta\phi} = 2 \ell^2 \sqrt{3  \chi_\kappa(\theta)} r,\qquad \mathcal{G}_{rt}=2 \ell q r,\label{TopMatConf1}
\end{equation}
provided that the cosmological constant $\Lambda$ and coupling $C$ are chosen as
\begin{equation}
	\Lambda = -\frac{z^2+z+4}{2\ell^2},\qquad \qquad C = \pm \frac{\sqrt{z}}{2 \sqrt{2} \ell}. \label{TopMatConf2}
\end{equation}
The function $\chi_\kappa(\theta)$ is
\begin{equation}
	\chi_\kappa(\theta)=\left\{\begin{array}{ccl}
		\sin \theta & \text { if } & \kappa=1, \\
		\theta & \text { if } & \kappa=0, \\
		\sinh \theta & \text { if } & \kappa=-1,
	\end{array}\right.
\end{equation}
For a line element of the form \eqref{lifshitzline}, this matter configuration yields the following metric function and Lifshitz exponent (cf.~\cite{Brynjolfsson:2009ct}; with $q$ normalized as in \eqref{TopMatConf1}, this $f$ solves the reduced equations for the above matter configuration):
\begin{equation}
	f(r) = 1+\frac{ \kappa }{10 r^2}-\frac{ 3 \kappa^2}{400 r^4}-\frac{q^2}{2 r^4},\qquad z=4,\label{ex1met}
\end{equation}
where $q$ is the electric charge.

Before turning to the double-copy map, we first examine the on-shell behavior of the reduced action. As noted above, the minisuperspace construction relies on an off-shell operator identity whose interpretation as a sourced Maxwell equation emerges only after imposing the Hamiltonian constraint.

For this solution the Hamiltonian constraint $\cLG+\cLN=0$ holds identically, so that the relation \eqref{DCdenkleme} requires no boundary term. Here $\cLM$ is understood with the electric term of the gauge field $\mathcal{G}$ in its Routhian form, while that of $\mathcal{F}$ is taken directly, since $\mathcal{F}$ is coupled to $\mathcal{B}$ through the $\mathcal{B}\wedge\mathcal{F}$ term. If this $\cLM$ is used directly, the sum $\cLG+\cLM$ does not vanish on shell but equals a total derivative,
\begin{equation}
	\eval{\cLG+\cLM}_{\text{on-shell}} = \frac{6}{5}\ell^2 \kappa r^3+\frac{3}{50}\ell^2 \kappa^2 r=-\partial_r \xi(r),
	\qquad
	\xi(r)=-\frac{3}{10}\,\ell^2\kappa\, r^4-\frac{3}{100}\,\ell^2\kappa^2 r^2 .
\end{equation}
The leading term, proportional to $\kappa r^4$, is of the same form as the gravitational boundary counterterm $\int \diff^{d-1}x \sqrt{\s}\, R(\s)\propto\kappa r^4$ built from the induced metric $\sigma_{ij}\diff x^i \diff x^j = \ell^2\left(-r^{2z} \diff t^2 + r^2 \diff \Sigma_{d-2, \kappa} \right)$ on a constant-$r$ hypersurface \cite{deHaro:2000vlm,Mann:2011hg,Ross:2011gu,Baggio:2011cp}. Since a total derivative does not contribute to the field equations, this difference is usually invisible in other double-copy formalisms; in the present formulation it does not affect the charge density, because the relation \eqref{DCdenkleme} is written in terms of $\cLN$.

The KS double copy of this solution follows from the above (cf.~\cite{Alkac:2025iyw}). The KS scalar and its electric field are
\begin{equation}
	\f_{KS}  = \left[-\frac{1}{10} \kappa  \ell^2 r^2+\frac{3}{400} \kappa ^2 \ell^2 +\frac{ \ell^2 q^2}{2}\right]r^4 \qquad\text{and}\qquad F_{rt} = \left[- \frac{3 \kappa \ell^2}{5}r^2 +\frac{3 \kappa^2 \ell^2}{100}   +2 \ell^2 q^2  \right]r^3,
\end{equation}
and the corresponding source is
\begin{equation}
	\rho_{KS} (r)=  -\frac{4 q^2}{\ell^2 r^4}-\frac{3 \kappa ^2}{50 \ell^2 r^4}+\frac{12 \kappa }{5 \ell^2 r^2} .
\end{equation}

On the other hand, the minisuperspace scalar is
\begin{align}
	\phi_{ms}(r) =& (\kappa-r^2)\ell^2 r^6+ \phi_{KS}\nn  \\
	=&\left[\frac{9}{10} \kappa  \ell^2 r^2+\frac{3}{400} \kappa ^2 \ell^2 +\frac{q^2}{2} -\ell^2 r^4\right] r^4,
\end{align}
and the single-copy potential $\bar\phi_{ms}=\phi_{ms}-\kappa\ell^2r^{6}=\phi_{KS}-\ell^2r^{8}$ is sourced by a minisuperspace charge of the form
\begin{align}
	\rho_{ms}(r) = & \rho_{KS}(r)+\Delta_{ms} \nn \\
	=& -\frac{4 q^2}{\ell^2 r^4}-\frac{3 \kappa ^2}{50 \ell^2 r^4}+\frac{12 \kappa }{5 \ell^2 r^2}+\frac{48}{\ell^2}\,.
\end{align}
The final constant term, $\frac{48}{\ell^2}$ in $\rho_{ms}$, comes from $\Delta_{ms}$ for $z=4$ and $d=4$.
	
	\subsection{Charged AdS Black Hole in Einstein--Gauss--Bonnet Gravity}

To further illustrate the robustness of the minisuperspace construction beyond two-derivative gravity, we briefly consider the charged AdS black hole solution in  Einstein--Gauss--Bonnet (EGB) gravity. Although this geometry is relativistic ($z=1$), it provides a useful test case for verifying that the minisuperspace operator identity remains consistent in the presence of higher-curvature corrections.

The choice of this example is motivated by the absence of known analytic Lifshitz black hole solutions in EGB gravity coupled to a Maxwell field. The Lovelock extension of Sec.~\ref{MDC} is therefore a structural statement that follows from the universal form of the reduced action; it applies to a Lifshitz solution of Lovelock gravity once one is found. We therefore test the minisuperspace construction in the relativistic limit $z=1$, corresponding to the AdS branch of the theory, where an exact charged black hole solution is available \cite{WILTSHIRE198636, PhysRevLett.55.2656, Cai:2001dz}.

We begin with the standard action of EGB theory coupled to a Maxwell field:
\begin{equation}
	S = \int \diff^{d}x \sqrt{-g} \left(R - 2\Lambda + \alpha \cG +\cLm \right) ,
\end{equation}
where
\begin{equation}
	\cG=R^2 - 4R_{\mu\nu}R^{\mu\nu} + R_{\mu\nu\rho\sigma}R^{\mu\nu\rho\sigma}
	\qquad\text{and}\qquad
	\cLm = -\frac{1}{4}f_{\m\n} f^{\m\n}.
\end{equation}
The Maxwell field strength tensor is defined as $f_{\m\n} =2 \partial_{[ \m } a_{ \n ]}$.

The matter configuration is taken as
\begin{equation}
	a_\m \diff x^\m = \vf(r) \diff t
	\qquad \text{and} \qquad
	\vf(r) = - q \frac{\sqrt{2(d-2) (d-3)}}{2\, r^{d-3}}
\end{equation}
and
\begin{equation}
	\Lambda = -\frac{(d-2) (d-1)}{2 \ell^2} ,
	\qquad
	\alpha = \frac{ \ell^2}{(d-3) (d-4)} \hat{\alpha}.
\end{equation}

For this matter configuration and our metric ansatz, given in \eqref{metricN} with the $z=1$ gauge choice and flat horizon ($\kappa=0$), the solution of the gravitational field equations is
\begin{equation}
	f(r) = \frac{1}{2 \hat{\alpha} } \left[  1 - \sqrt{1 -  4 \hat{\alpha} \left(1 - \frac{\mu }{r^{d-1}} + \frac{q^2}{\ell^2 r^{2d-4}}\right) }  \right].
\end{equation}

We now proceed with the minisuperspace double-copy analysis. The minisuperspace Wheeler polynomial $\mathcal{W}_{ms}$ takes the form
\begin{equation}
	\mathcal{W}_{ms} = \ell^2 r^2 \left(-f+\hat{\alpha} f^2\right).
\end{equation}

Using the solution above, $\hat\alpha f^2-f=-\big(1-\mu/r^{d-1}+q^2/(\ell^2r^{2d-4})\big)$, so that $\mathcal{W}_{ms}$ is independent of $\hat\alpha$; this is expected, since the Hamiltonian constraint fixes $\mathcal{W}_{ms}$ entirely in terms of the matter content. According to the discussion in Sec.~\ref{MDC}, the Wheeler minisuperspace polynomial $\mathcal{W}_{ms}$ acts as an effective electric potential in the minisuperspace double-copy formalism and is associated with a charge distribution
\begin{equation}
	\rho_{ms} = \rho_{KS} + \Delta_{ms}.
\end{equation}

Applying the radial Maxwell operator of Eq.~\eqref{msMaxwell} to $\mathcal{W}_{ms}$, and denoting by $\rho_{KS}\equiv\rho_{ms}-\Delta_{ms}$ the contribution of the gauge field, we find
\begin{equation}
	\rho_{KS} = \frac{2(d-3)^2 q^2}{\ell^4 r^{2d-4}},
	\qquad
	\Delta_{ms} = \frac{2(d-1)}{\ell^2}.
\end{equation}

Thus, even in the presence of Gauss--Bonnet higher-curvature corrections, the minisuperspace Wheeler polynomial continues to act as an effective electric potential whose source structure is captured by $\rho_{KS}$ and the constant shift $\Delta_{ms}$.

	\section{Conclusion} \label{Conc}
	
	In this work, we developed a minisuperspace realization of the classical double copy for anisotropic Lifshitz spacetimes. By imposing static symmetries directly at the level of the action, we obtained a universal reduced gravitational Lagrangian of total-derivative form, with all model dependence encoded in a single minisuperspace potential. We then introduced a weighted radial operator $\Dz$, fixed by Lifshitz scaling and radial diffeomorphism invariance, and established an off-shell operator identity: acting with $\Dz$ on the reduced gravitational Lagrangian reproduces the canonical Maxwell operator for the temporal single-copy potential. In this formulation, the Maxwell structure appears as a kinematical property of the reduced system, rather than as a consequence of a particular solution ansatz.
	
	Away from maximal symmetry, the identity is supplemented by a residual deviation term $\Zms(r)$. We showed that $\Zms$ is not an arbitrary remainder, but is fixed by Lifshitz anisotropy together with the intrinsic curvature of the horizon sections. Its structure separates into a potential-dependent contribution controlled by the universal prefactor $(z-1)(d-2z-1)$ and a term linear in $\kappa$ with $r^{-2}$ scaling that removes the horizon-curvature piece from the single-copy potential. Both contributions vanish smoothly in the relativistic limit $z\to 1$, recovering the AdS case in which no additional deviation term appears beyond the cosmological constant.
	
	Imposing the radial Hamiltonian constraint provides the on-shell completion of this identity, with the matter sector generating the effective Maxwell source. In regimes where a Kerr--Schild double copy description exists, the sourced Maxwell equation derived from the reduced constraint reproduces the corresponding Kerr--Schild single-copy charge density, together with an additional constant contribution associated with non-maximal symmetry when $\Lambda$ is treated within the reduced matter sector. This recasts the classical double copy as a property of the radial constraint structure, rather than as a feature tied solely to algebraically special metric decompositions.
	
	We further showed that the same mechanism persists in Lovelock theories. Replacing the Einstein minisuperspace potential by the Kerr--Schild--weighted Wheeler polynomial preserves the off-shell operator identity and yields the same universal form of the deviation term. This robustness indicates that the minisuperspace double-copy structure is rooted in symmetry-reduced gravitational dynamics and remains stable under higher-curvature deformations that preserve second-order field equations.
	
	Explicit Lifshitz black hole solutions illustrate how matter content and horizon topology enter the sourced Maxwell equation in the reduced framework. These examples include cases with vacuum single copy, nontrivial charge distributions, and topological contributions, for which the constraint holds identically in terms of $\cLN$ while $\cLM$ differs from it by a total derivative related to boundary counterterms. Together, they demonstrate that the minisuperspace construction cleanly separates an off-shell geometric identity from its on-shell sourced completion, while maintaining control over deviations from maximal symmetry.
	
	As an additional consistency check, we examined the charged AdS black hole solution in Einstein--Gauss--Bonnet gravity. Although this geometry corresponds to the relativistic limit $z=1$, it provides a useful test of the minisuperspace operator identity in the presence of higher-curvature corrections. We found that the Wheeler minisuperspace polynomial continues to act as an effective electric potential, with a source structure determined by the Kerr--Schild contribution together with a universal background term.
	
	These results suggest that the minisuperspace formulation captures a structural aspect of the classical double copy that extends beyond two-derivative gravity and remains robust under both anisotropic scaling and higher-curvature deformations.
	
	Several directions for future work remain. A natural extension is to relax the static ansatz considered here and investigate rotating or time-dependent configurations depending only on the radial coordinate, where the reduced Hamiltonian structure becomes richer and additional components of the Maxwell operator may emerge. It would also be interesting to explore whether similar operator identities persist in higher-curvature theories beyond Lovelock gravity, where the constraint structure is modified. Another natural direction is to formulate the zeroth-copy equation within the reduced framework, which is not addressed here.
	
	From a holographic perspective, the minisuperspace formulation may provide a useful viewpoint for Lifshitz holography. Since the radial coordinate tracks scaling behavior in the reduced system, the operator $\Dz$ organizes the dynamics according to Lifshitz weights and may offer a geometric interpretation of renormalization-group–like flow in the dual field theory.
	
	More broadly, these results suggest that part of the classical double-copy structure may be encoded directly in the constraint algebra of symmetry-reduced gravity, providing a complementary perspective to Kerr--Schild double copy and Weyl double copy constructions.
	
	\section*{Acknowledgments}

The author is grateful to Dr. G\"okhan Alka\c{c} for valuable discussions.

\section*{Declaration of Generative AI and AI-assisted technologies in the writing process}

During the preparation of this work the author used Claude Sonnet 5.5 (Anthropic) in order to improve the language and readability of the manuscript. After using this tool, the author reviewed and edited the content as needed and takes full responsibility for the content of the publication.

\appendix

\section{Contraction of $E^{\mu}{}_{\nu}$ with a Killing vector}
\label{app:Killing}

The cancellation of the extra term in Sec.~2 has a simple geometric origin.
To see this, we evaluate $E^{\mu}{}_{\nu}V^{\nu}$ without using the field
equations. We take $k^\mu$ to be affinely parametrized, as is implicit in
Eq.~\eqref{etensor}, and assume that $\phi_{\rm KS}$ and $k^\mu$ share the
symmetry generated by $V^\mu$, namely
$\mathcal{L}_V\phi_{\rm KS}=0$ and $\mathcal{L}_Vk^\mu=0$, where
$\mathcal{L}_V$ is the Lie derivative along $V^\mu$. Unless the full metric appears explicitly, indices are
raised and lowered with the background metric.

\subsection{General identity}

Instead of expanding Eq.~\eqref{etensor}, it is simpler to contract
Eq.~\eqref{riccif} with $V^\nu$, which gives
\begin{equation}
E^{\mu}{}_{\nu}V^{\nu}
=-2\left(R^{\mu}{}_{\nu}-\bar R^{\mu}{}_{\nu}\right)V^{\nu}
-(V\cdot k)\,\bar\nabla_\alpha F^{\alpha\mu}.
\label{eq:A-start}
\end{equation}
It is therefore sufficient to compute the difference of the two Ricci
tensors along $V^\nu$. For a Killing vector $V^\mu$ of the full metric, the Ricci tensor satisfies
\begin{equation}
R^{\mu}{}_{\nu}V^{\nu}=\nabla_\nu\nabla^\mu V^\nu ,
\label{eq:A-Killing}
\end{equation}
where the covariant derivatives and the raised index on the right-hand side
refer to the full metric. By the Killing equation, $\nabla^\mu V^\nu$ is
antisymmetric. Its divergence can therefore be written in terms of partial
derivatives as
\begin{equation}
R^{\mu}{}_{\nu}V^{\nu}
=\frac{1}{2\sqrt{-g}}\,\partial_\nu\!\left\{\sqrt{-g}\,
g^{\mu\alpha}g^{\nu\beta}
\left[\partial_\alpha\!\left(g_{\beta\rho}V^\rho\right)
-\partial_\beta\!\left(g_{\alpha\rho}V^\rho\right)\right]\right\}.
\label{eq:A-explicit}
\end{equation}
The expression in square brackets is the only place where the covariant
components of the Killing vector enter. It is therefore convenient to define
\begin{equation}
\Upsilon_{\alpha\beta}\equiv
\partial_\alpha\!\left(g_{\beta\rho}V^\rho\right)
-\partial_\beta\!\left(g_{\alpha\rho}V^\rho\right).
\label{eq:A-def}
\end{equation}
Equation~\eqref{eq:A-explicit} also holds for the background, where the
corresponding tensor is
$\bar\Upsilon_{\mu\nu}=\partial_\mu \bar V_\nu-\partial_\nu \bar V_\mu$,
with $\bar V_\mu\equiv\bar g_{\mu\nu}V^\nu$. Since
$\sqrt{-g}=\sqrt{-\bar g}$, subtracting the two expressions gives
\begin{equation}
2\left(R^{\mu}{}_{\nu}-\bar R^{\mu}{}_{\nu}\right)V^{\nu}
=\bar\nabla_\nu\!\left(
g^{\mu\alpha}g^{\nu\beta}\,\Upsilon_{\alpha\beta}
-\bar\Upsilon^{\mu\nu}\right).
\label{eq:A-Ricci}
\end{equation}

The contravariant components $V^\mu$ are the same for both metrics, whereas
$g_{\nu\alpha}V^\alpha=\bar V_\nu+(V\cdot k)A_\nu$. Consequently,
\begin{equation}
\Upsilon_{\mu\nu}=\bar\Upsilon_{\mu\nu}
+\partial_\mu\!\left[(V\cdot k)A_\nu\right]
-\partial_\nu\!\left[(V\cdot k)A_\mu\right].
\label{eq:A-lower}
\end{equation}
Using $g^{\mu\nu}=\bar g^{\mu\nu}-\phi_{\rm KS}k^\mu k^\nu$ and
$k^\alpha k^\beta\Upsilon_{\alpha\beta}=0$, we also find
\begin{equation}
g^{\mu\alpha}g^{\nu\beta}\,\Upsilon_{\alpha\beta}
=\bar g^{\mu\alpha}\bar g^{\nu\beta}\,\Upsilon_{\alpha\beta}
-\phi_{\rm KS}\,k^\alpha\,\Upsilon_{\alpha\beta}
\left(k^\mu\bar g^{\nu\beta}-k^\nu\bar g^{\mu\beta}\right).
\label{eq:A-inverse}
\end{equation}
To evaluate $k^\alpha\Upsilon_{\alpha\beta}$, we treat the two parts of
Eq.~\eqref{eq:A-lower} separately. For the background part, the Killing
equation $\bar\nabla_\alpha \bar V_\beta=-\bar\nabla_\beta \bar V_\alpha$ and the
condition $\mathcal{L}_Vk_\beta=V^\alpha\bar\nabla_\alpha k_\beta
+k_\alpha\bar\nabla_\beta V^\alpha=0$ give
\begin{equation}
k^\alpha\bar\Upsilon_{\alpha\beta}
=-2k^\alpha\bar\nabla_\beta \bar V_\alpha
=2V^\alpha\bar\nabla_\alpha k_\beta .
\end{equation}
For the remaining part, $k^2=0$ implies
$k^\alpha\bar\nabla_\beta k_\alpha=0$. Together with the geodesic equation,
this gives
\begin{equation}
k^\alpha\left\{\partial_\alpha\!\left[(V\cdot k)A_\beta\right]
-\partial_\beta\!\left[(V\cdot k)A_\alpha\right]\right\}
=k^\alpha\partial_\alpha\!\left[(V\cdot k)\,\phi_{\rm KS}\right]k_\beta .
\end{equation}
Therefore,
\begin{equation}
k^\alpha\Upsilon_{\alpha\beta}
=2V^\alpha\bar\nabla_\alpha k_\beta
+k^\alpha\partial_\alpha\!\left[(V\cdot k)\,\phi_{\rm KS}\right]k_\beta .
\label{eq:A-kUpsilon}
\end{equation}
The term proportional to $k_\beta$ drops out of Eq.~\eqref{eq:A-inverse}.
On the other hand, raising the indices in Eq.~\eqref{eq:A-lower} with the
background metric yields
\begin{equation}
\bar g^{\mu\alpha}\bar g^{\nu\beta}\,\Upsilon_{\alpha\beta}
=\bar\Upsilon^{\mu\nu}+(V\cdot k)F^{\mu\nu}
+\phi_{\rm KS}\left[k^\nu\partial^\mu(V\cdot k)
-k^\mu\partial^\nu(V\cdot k)\right].
\end{equation}
Collecting these results, we obtain
\begin{equation}
\begin{split}
g^{\mu\alpha}g^{\nu\beta}\,\Upsilon_{\alpha\beta}
-\bar\Upsilon^{\mu\nu}
=(V\cdot k)F^{\mu\nu}
+\phi_{\rm KS}\big[
&\left(\partial^\mu(V\cdot k)+2V^\alpha\bar\nabla_\alpha k^\mu\right)k^\nu\\
{}-{}&\left(\partial^\nu(V\cdot k)+2V^\alpha\bar\nabla_\alpha k^\nu\right)k^\mu
\big].
\end{split}
\label{eq:A-difference}
\end{equation}
The same combination appears in both terms inside the square brackets. It is
therefore natural to define
\begin{equation}
w^\mu\equiv\partial^\mu(V\cdot k)+2V^\alpha\bar\nabla_\alpha k^\mu .
\label{eq:A-wdef}
\end{equation}
Since
$\mathcal{L}_Vk_\mu=V^\alpha\left(\partial_\alpha k_\mu
-\partial_\mu k_\alpha\right)+\partial_\mu(V\cdot k)=0$, the corresponding covector
can also be written as
\begin{equation}
w_\mu=V^\alpha\left(\bar\nabla_\alpha k_\mu
+\bar\nabla_\mu k_\alpha\right)
=V^\alpha\left(\mathcal{L}_k\bar g\right)_{\alpha\mu}.
\label{eq:A-w}
\end{equation}
We can now return to Eq.~\eqref{eq:A-Ricci}. We use
Eq.~\eqref{eq:A-difference}, expressed in terms of $w^\mu$, together with the
antisymmetry of $F^{\mu\nu}$, and multiply the result by $-1$. This gives
\begin{equation}
\begin{split}
-2\left(R^{\mu}{}_{\nu}-\bar R^{\mu}{}_{\nu}\right)V^{\nu}
={}&(V\cdot k)\,\bar\nabla_\alpha F^{\alpha\mu}
+F^{\alpha\mu}\,\partial_\alpha(V\cdot k)\\
&+\bar\nabla_\alpha\!\left[\phi_{\rm KS}
\left(w^\alpha k^\mu-w^\mu k^\alpha\right)\right].
\end{split}
\label{eq:A-Riccifinal}
\end{equation}
The left-hand side is precisely the first term on the right-hand side of
Eq.~\eqref{eq:A-start}. Upon substitution, the two terms containing
$(V\cdot k)\,\bar\nabla_\alpha F^{\alpha\mu}$ cancel each other, and we
arrive at
\begin{equation}
E^{\mu}{}_{\nu}V^{\nu}
=F^{\alpha\mu}\,\partial_\alpha(V\cdot k)
+\bar\nabla_\alpha\!\left[\phi_{\rm KS}
\left(w^\alpha k^\mu-w^\mu k^\alpha\right)\right].
\label{eq:A-master}
\end{equation}
This identity holds under the conditions stated above. In general, the right-hand
side does not vanish.

\subsection{Sufficient conditions for $E^{\mu}{}_{\nu}V^{\nu}=0$}

Suppose that $V\cdot k$ is constant. Then $\mathcal{L}_Vk_\mu=0$ gives
$w_\mu=2V^\alpha\bar\nabla_\alpha k_\mu$. Using $\mathcal{L}_Vk^\mu=0$ once
more,
\begin{equation}
w^\mu=2V^\alpha\bar\nabla_\alpha k^\mu
=2k^\alpha\bar\nabla_\alpha V^\mu .
\end{equation}
It then follows from Eq.~\eqref{eq:A-master} that
\begin{equation}
V\cdot k=\text{constant}
\quad\text{and}\quad
k^\alpha\bar\nabla_\alpha V^\mu=\lambda\,k^\mu
\qquad\Longrightarrow\qquad
E^{\mu}{}_{\nu}V^{\nu}=0,
\label{eq:A-condition}
\end{equation}
where $\lambda$ is a scalar function. In other words, it suffices that
$k^\mu$ be an eigenvector of the Killing two-form
$\bar\nabla_\mu \bar V_\nu$. This can be seen directly from
Eq.~\eqref{eq:A-master}. Its first term vanishes because $V\cdot k$ is
constant. Moreover, the eigenvalue equation implies
$w^\mu=2\lambda\,k^\mu$, so that the combination appearing in its second
term becomes
\begin{equation}
w^\alpha k^\mu-w^\mu k^\alpha
=2\lambda\left(k^\alpha k^\mu-k^\mu k^\alpha\right)
=0 .
\label{eq:A-bracket}
\end{equation}
The second term therefore vanishes identically, and the explicit form of
$\lambda$ plays no role. These conditions are sufficient; we do not claim that they are necessary.
We also note that they refer to a given pair $(\phi_{\rm KS},k^\mu)$. A
rescaling $k^\mu\to e^{\sigma}k^\mu$,
$\phi_{\rm KS}\to e^{-2\sigma}\phi_{\rm KS}$ leaves the metric unchanged,
but in general changes $A^\mu$ and hence $E^{\mu}{}_{\nu}$. In particular,
setting $V\cdot k=1$ by such a rescaling is not by itself sufficient.

\subsection{Static backgrounds}

Consider the background
\begin{equation}
d\bar s^2=-h(r)\,dt^2+\frac{dr^2}{p(r)}+C(r)\,d\Sigma^2_{d-2,\kappa},
\end{equation}
with $h,p>0$, and take
\begin{equation}
k_\mu dx^\mu=dt+\frac{dr}{\sqrt{h\,p}},
\qquad
V=\partial_t,
\qquad
\phi_{\rm KS}=\phi_{\rm KS}(r).
\end{equation}
Here $k_\mu$ is null and is the gradient of a scalar, so it is affinely
geodesic. Moreover, $V\cdot k=1$. The only independent nonvanishing
component of $\bar\nabla_\mu \bar V_\nu$ is the $tr$ component. Hence the vector
$k^\alpha\bar\nabla_\alpha V^\mu$ lies in the $(t,r)$ plane, and by the
antisymmetry of $\bar\nabla_\alpha \bar V_\mu$ it is orthogonal to $k^\mu$. In a
two-dimensional Lorentzian plane, a vector orthogonal to a null vector is
proportional to it. The conditions in Eq.~\eqref{eq:A-condition} are
therefore satisfied identically, with
\begin{equation}
\lambda=-\frac{h'}{2}\sqrt{\frac{p}{h}} .
\end{equation}
For the Lifshitz background, $h=\ell^2r^{2z}$, $p=r^2/\ell^2$, and
$C=\ell^2r^2$, one finds $\lambda=-z\,r^{z}$. Note that $\lambda$ does not vanish for
$z=1$ either. In particular, for $\kappa=0$ the background then reduces to
maximally symmetric AdS spacetime, for which $\lambda=-r$. A nonvanishing
$\lambda$ is therefore not a consequence of the Lifshitz anisotropy.

 We conclude that
\begin{equation}
E^{\mu}{}_{\nu}V^{\nu}=0
\end{equation}
for any $d$, $z$, $\kappa$, and $f(r)$ in the static Lifshitz ansatz,
whether or not $f(r)$ solves the field equations. It follows that
$E^\mu=0$ and $\mathcal{E}=0$. Thus $V^\nu$ is a right eigenvector of
$E^{\mu}{}_{\nu}$ with zero eigenvalue, and consequently
$\det(E^{\mu}{}_{\nu})=0$.

\bibliographystyle{apsrev4-1}
\bibliography{references}
	
\end{document}